\documentclass[12pt]{article}
\usepackage{graphicx} 
\usepackage{amsmath}
\usepackage{amssymb}
\usepackage{cite}
\usepackage{geometry}

\usepackage{mathrsfs}
\usepackage{authblk}
\usepackage{breqn}
\usepackage{subcaption}
\usepackage{float}
\usepackage[pagebackref=false, colorlinks=true]{hyperref}
\hypersetup{
	linkcolor=blue,     
	citecolor=blue,     
	urlcolor=blue}
\usepackage{xcolor}

\definecolor{myred}{HTML}{C44E52}
\definecolor{myblue}{HTML}{4C72B0}
\definecolor{mygreen}{HTML}{55A868}
\definecolor{mypurple}{HTML}{8172B3}
\definecolor{myorange}{HTML}{DD8452}

\title{\boldmath Stability of Vacuum and Quantum Dynamics in Anti-de Sitter Spacetimes}
\author[1]{Shi-Yuan Li,\thanks{lishy@sdu.edu.cn}}
\affil[1]{School of Physics, Shandong University, Jinan, Shandong, 250100, P.R.China}
\author[1]{Chengwu Liu\thanks{liuchengwu@mail.sdu.edu.cn}}

\date{\today}

\begin{document}
	
	\maketitle
	\begin{abstract}
		Canonical quantization on scalar and Maxwell fields is studied for the stability of vacuum and construction of quantum dynamics in Anti-de Sitter spacetimes.
		The $\xi R\phi^2$ term in the Lagrangian of scalar field with $\xi > 5/48$ (mass term just introduces shift of $\xi$) can 
		cause  the   spectrum of the self-adjoint  Hamiltonian  with  negative  value    not bounded from below. The  degrees of freedom corresponding to the negative values  are  constrained system and are ambiguous  in space  direction and in commutation relation.   The only other algebra accommodated in 3-dimensional space consistent with locality,     the anti-commutation relation, is definite confronting the constrain, and trivialize the     negative  spectrum to be inactive and  absent degrees of freedom.     Thus only the physical bosons corresponding to the non-negative spectrum span the Hilbert space. 
		For the Maxwell fields as another constrained system, the gauge invariant  canonical energy momentum tensor   straightforwardly leads to the gauge invariant non-negative Hamiltonian. The 2-dimensional independent dynamical degrees of freedom are quantized in temporal gauge.
		The energy momentum tensors for both quantized fields are renormalized to be finite at operator level, which guarantee  the stable vacuum maximally symmetric and the  particle Fock states the representation of the AdS symmetry SO(2,3).
		Adopting  Weinberg's framework, the  new covariant free quantum scalar  field is reconstructed in  the corresponding mode function space  isomorphic to the free boson Hilbert space,  allowing   various local effective interaction terms among   fields  introduced. The Yang-Feldman equations  relate the free fields and the interacting fields with the   Green functions determined  properly in the mode function space.
		
	\end{abstract}
	\section{Introduction}
	The intrinsic difficulties in constructing the 'completely quantum theory' of all the matters as well as the gravity (spacetime)  indicate that deeper physical principles or deeper understanding on the present quantum principles  may be needed.
	On the other hand, a rather {\it modest} way out for deeper understanding on the quantum principles can be the intensive studies on various effective quantum field theories (EQFT) including  the spacetime perturbations  in various backgrounds of curved spacetimes, and reversely on the  backreactions. 
	Such studies can also  shed light on  dilemma in Minkowski spacetime (a recent example on dilemma from infinite momentum integrations, see \cite{Li:2023aex}), and on
	quantum fields interacting with background fields, a practically  important topic in optics and condensate matter sustaining from the  early time \cite{Furry:1951bef}. The general framework of algebraic quantum field theory in curved spacetime with causal structure has long been setup, see, e.g.,  \cite{Kay:2023vbi, Fewster:2015kua, Fredenhagen:2014lda, Hollands:2014eia, Benini:2013fia} for recent reviews. There are several special cases of curved spacetime need much more concrete studies in details, such as our cosmic spacetime and the Anti-de Sitter (AdS) spacetime, since for both cases the EQFT and the quantum principles can confront various tests via their predictions relevant to the experimental measurements. 
	
	EQFT in AdS gives predictions for physical observables with the help of the AdS/CFT correspondence \cite{Maldacena:1997re, Gubser:1998bc, Witten:1998qj}, based on which the (in)stability of the AdS confronting the  perturbations from the matter fields is a key question. The AdS instability conjecture by  Dafermos and Holzegel in 2006,\cite{Dafermos, Dafermos2} states that
	there exist arbitrarily small perturbations to AdS initial data which, under evolution by the Einstein vacuum
	equations for $\Lambda <0$  with reflecting boundary conditions on conformal infinity, lead to the formation of black
	holes.  The numerical study of this conjecture in the simpler setting of the spherically symmetric Einstein-scalar
	field system was initiated by Bizon and Rostworowski \cite{PB}, followed by a vast number of numerical and heuristic
	works by several authors. At present, cases such as the classical Einstein-scalar \cite{PB, JR, DF}, pure-gravity \cite{DC} Einstein-Vlassov system \cite{G2018}, Einstein-null dust \cite{G2017} have been numerically and analytically studied. 
	The seek for rigorous study with simplest dynamics \cite{Moschidis:2018kcf, G2018, G2017} is with  non-trivial boundary condition. 
	On the other hand, there are also stable solutions which are named as islands of stability \cite{GS, MR} such as geons \cite{DC2} for pure gravity system and \cite{MM} for Einstein massless scalar system. 
	At the same time, back-reaction in the framework of quantized scalar field is also studied \cite{JCT, Thompson:2025jkn} with some special boundary conditions.
	
	The investigations relevant to the (in)stability of the AdS and AdS/CFT highlight the need for various concrete EQFT in AdS \cite{AS, Wald2, Wald, AC1, Bertan:2018afl} as working models for the extension of the present recognitions. 
	Canonical quantization \cite{Parker, Fulling:1989nb} provides a promising way for setting up a self-consistent effective dynamics, especially it places the stability of the quantum vacuum under control. 
	In this paper, this framework is adopted for the non-minimal coupled real scalar fields and Maxwell field. It is found that these two quantum dynamics can both have the stable vacuum state consistent with the background AdS geometry, and have the global Hilbert space defined by the physical particle Fock states, even for the value of $\xi$ with classical instability. 
	The AdS metric is time-independent so that a global Hilbert space possibly exists  \cite{Wald, Bertan:2018afl}. This is different from the cosmic spacetime, of which the metric and hence the Hamiltonian of a quantum system are explicitly time dependent and in general the global (in time) unique Hilbert space for such quantum system can not be defined; and special Hadamard states are employed to link the global curved cosmic spacetime EQFT and the local Minkowski, where the measurement is performed.


	Particularly, the Hamiltonian is arranged in the diagonal form (\`{a} la Bogoliubov), 
	from which one can easily sees whether the Hamiltonian is bounded from below and whether the vacuum state (ground state) can be defined.
	For the massless case, if $\xi\le 5/48$, the diagonalized Hamiltonian is non-negative and the vacuum and particle Fock states are well defined. However, if $\xi> 5/48$ the Hamiltonian can have negative spectrum not bounded from below. It follows that the negative sector is a constrained system having uncertain direction of angular-momentum and indefinite commutation relation. The only possible algebraic anti-commutation relation is employed, which trivializes the negative energy spectrum and obtains the minimum of the Hamiltonian for the definition of the vacuum. After the above quantization, the renormalization of the canonical energy momentum tensor operator $T^{\mu\nu}$ is investigated. We take the BPHZ-like 'maximum' subtraction scheme, to obtain $T_R^{\mu\nu}$ finite the quantum vacuum maximally symmetric. 
	The renormalized energy momentum tensor automatically gives the correct subtraction term of the Hamiltonian with the help of the time-like Killing vector. These results further guarantee the negative degrees of freedom absent and decoupled from the dynamics. Thus for any value of $\xi$ a Hilbert space can be well defined by the physical particle Fock states. Following the framework of Weinberg\cite{weinberg1995quantum}, a new free field $\chi$ and interacting field $\tilde{\phi}$ with various effective interaction Lagrangians can be constructed. The relation between $\tilde{\phi}$ and $\chi$ is described by the Yang-Feldman equations. This construction intrinsically covers the causality, unitarity and locality. 
	All the above investigations  for real scalar field are contents of Section 2.

	In section 3, we study the canonical quantization  of  the electromagnetic field, and demonstrate that 
	the  gauge invariance plays the key r\^ole. 
	It has already been recognized in Minkowski spacetime and with the Coulomb gauge  that  the independent degrees of freedom are not necessarily  the correct canonical degrees of freedom which lead to the diagonalized Hamiltonian.
	Since the canonical  $T^{\mu\nu}$ and the corresponding Hamiltonian obtained via the help of the time-like Killing vector are both gauge invariant,
	we adopt the  the expansion of the Maxwell filed in AdS given by the  Refs.   \cite{Ruffini, Wh, CV} and taking the temporal gauge.  
	Two dimensions of dynamical  degrees of freedom are canonically quantized.
	The vacuum of Maxwell field is stable since the gauge invariant Hamiltonian is definitely non-negative.
	The renormalization  
	are similar as those of the scalar field. And various effective interaction terms can be introduced as section 2. 
	
	The last section, section 4, is for conclusions and discussions, which include some open questions of
	the renormalization and the consistent study of the backreaction from the EQFT source to the background geometry,  and the  interaction between fields.

	\par\vskip 1cm\noindent

	As is well known, though the algebraic quantum field theory takes the  covariant form, the Hilbert space of states is not unique (not  equivalent via a unitary transformation) under the general spacetime transformations.
	So we can only make use of the canonical quantization framework relying on the specific parameterization of the AdS as listed in the following.
	Needless to say, only by investigating all  inequivalent Hilbert spaces of states,  one may get a thorough understanding of the covariant quantum principles.

	In the presence of a negative cosmological constant $\Lambda$, the maximally symmetric solution of the vacuum Einstein equations
	\begin{equation}
		\label{eeq}
		R_{\mu\nu}-\frac{1}{2}Rg_{\mu\nu}+\Lambda g_{\mu\nu}=0
	\end{equation}
	in $n+1$ dimensions, is the  $AdS_{n+1}$ spacetime. 
	The     (covering) AdS  metric employed  in this paper (taking the example $n=3$) is
	\begin{equation}
		\label{lemetric}
		g_{\mu\nu}=\mathrm{diag}\left\lbrace -\left(1+l^{-2}r^2\right),\left(1+l^{-2}r^2\right)^{-1},r^2\gamma_{ij}\right\rbrace.
	\end{equation}
	$i=1,2$ is the coordinate chart on $\mathbb{S}^{2}$ and $\Lambda=-3l^{-2}$ with $l^{-2}$ the AdS radius.
	
	The Lagrangian density of the real scalar field  we study in this paper is
	\begin{equation}\label{lagrangian}
		\mathcal{L}=\sqrt{-g}\frac{1}{2}\left(-g^{\mu\nu}\partial_\mu\phi \partial_\nu \phi -m^2 \phi^2 -\xi R\phi^2\right),
	\end{equation}
	with  the coupling to the curvature  taken into account;
	and that of the Maxwell field is
	\begin{equation}\label{MF}
		\mathcal{L}=-\frac{1}{4}\sqrt{-g}F_{\mu\nu} F^{\mu\nu},~~~~~\:F_{\mu\nu}=\partial_\mu A_\nu -\partial_\nu A_\mu.
	\end{equation}

	\section{Scalar Field}
	Since the scalar curvature of AdS is a constant, the mass term in the Lagrangian density is just a shift of $\xi$ with $\xi \to \xi + \frac{m^2}{R}$. Here we mainly consider the massless case, the mass can be trivially taken into account via the shift of $\xi$.
	To proceed the canonical quantization we employ the standard way of Legendre transformation to obtain the
	conjugate momentum of $\phi$,
	\begin{equation}\label{conjugate momentum}
		\Pi=\frac{\delta L}{\delta \dot{\phi}}=\frac{r^2\sin\theta}{1+l^{-2}r^2}\dot{\phi},
	\end{equation}
	with $L$ the spacial integration of the Lagrangian density (Hereafter in this paper overdots or primes denote derivative with respect to $t$ or $r$ respectively),
	and the  Hamiltonian
	\begin{align}
		H&=\int dr d\theta d\varphi \left(\Pi \dot{\phi}-\mathcal{L}\right) \\
		\label{H21}&=\int drd\theta d\varphi\frac{r^2\sin\theta}{2}\left(\frac{1}{1+l^{-2}r^2}\dot{\phi}^2+(1+l^{-2}r^2)\phi^{\prime 2}+\frac{1}{r^2}\partial_\theta\phi\partial_\theta\phi+\frac{1}{r^2\sin^2\theta}\partial_\varphi\phi\partial_\varphi\phi+\xi R \phi^2\right)\\
		\label{H2}&=\frac{1}{2}\int dr d\theta d\varphi \frac{r^2\sin\theta}{1+l^{-2}r^2}\left(\dot{\phi}^2-\ddot{\phi}\phi\right).
	\end{align}
	From Eq. \eqref{H21} to Eq. \eqref{H2}, the equations of motion and the  natural periodic boundary conditions for $\theta$ and $\varphi$ have been used, leaving a term corresponding to the boundary of the radial boundary
	\begin{equation}\label{ST}
		\frac{1}{2}\int dr d\theta d\varphi \partial_r\left(\sin\theta r^2(1+l^{-2}r^2)\phi \phi^\prime\right), 
	\end{equation}
	which only depends on radial boundary conditions of $\phi, \phi^\prime$, this term will be investaged for various cases in the following. 
	
	The Hamiltonian is not positive definite since the scalar curvature of AdS is negative. In Eq. \eqref{H21}, the last term is negative for $\xi >0$, and all the other terms are positive.
	How the spectrum of the quantized Hamiltonian  is dependent  on the  value of $\xi$,  especially, the range of  $\xi$ in which the spectrum of the Hamiltonian is bounded from below and the vacuum is well-defined can not be straightforwardly read out from the
	above form of the Hamiltonian. 	For the answer we need to study the details of the quantization, and we will see the result is not  $\xi \le 0$.
	To  employ the  canonical quantization formulations 	the key point is to
	find the proper canonical observables and the corresponding canonical commutation relations, with which the Hamiltonian is expressed and the spectrum is obtained.
	For this purpose, we should not \`{a} priori insert the canonical commutative relations on $\phi$ and $\Pi$, but first to explore the mode functions
	on which the field can be expanded. From this expansion, and if the concrete form of the Hamiltonian can be diagonalized \cite{bogolubov2010introduction}, we can determine the canonical variables, the commutation relations, as well as the spectrum. 
	This wisdom has been employed in the quantization of  gauge theory in Minkowski spacetime (e.g., quantum electrodynamics in Coulomb gauge) and 
	will also be applied in next section  for  Maxwell field with the  independent degree of freedom singled out. 
	
	It is easy to get the equation of motion of $\phi$ from \eqref{lagrangian} 
	\begin{equation}\label{F}
		\hat{F}[\phi]=\left[g^{\mu\nu}\left(\partial_\mu \partial_\nu-\Gamma^\rho_{\mu\nu}\partial_\rho\right)-\xi R\right]\phi=0. 
	\end{equation}
	A solution of the Eq. \eqref{F} can be expressed as 
	\begin{equation}\label{10}
		\phi_{\lambda j m}(t,r,\theta,\varphi)=\phi_\lambda(t)u_{\lambda j}(r)Y_{jm}(\theta,\varphi), 
	\end{equation}
	via the separation of variables, where $Y_{jm}(\theta,\varphi)$ is the spherical harmonic function, as the result of the spatial symmetry. From the classcial equation of motion
	\begin{align}
		\label{eom1}&\ddot{\phi}_\lambda(t)=-\lambda\phi_\lambda(t), \\
		\label{eom}&A ~u_{\lambda j}=\frac{\lambda r^2}{1+l^{-2}r^2}u_{\lambda j},
	\end{align}
	with the operator $A$ defined as\cite{Wald} 
	\begin{equation}\label{A}
		A~u_{\lambda j}\equiv -\left(r^2(1+l^{-2}r^2)u_{\lambda j}^\prime\right)^\prime+j(j+1)u_{\lambda j}+\xi Rr^2u_{\lambda j}. 
	\end{equation}
	Hence for the real scalar field in AdS, the mode functions are completely determined when the solution of Eq. \eqref{eom} are obtained.
	The spectrum of the  Hamiltonian is also determined by the spectrum of the operator $A$, Eq. \eqref{A}. 
	Thanks to Wald\cite{Wald}, the eigenfunctions and spectrum of the symmetric operator $A$ have been completely investigated.  The self-adjoint extension of $A$ can be obtained corresponding to a certain radial boundary condition, which guarantees the dimensions of the deficiency subspaces $n_{\pm}$ equals to each other, and $u_{\lambda j}$ is orthonormal eigenfunctions of the self-adjoint operator $A$, i.e. 
	\begin{equation}\label{1}
		\int_0^\infty \frac{u_{\omega j}u_{\omega^\prime j}r^2dr}{1+l^{-2}r^2}=\frac{\delta(\omega-\omega^\prime)}{l}, 
	\end{equation}
	where $u_{\lambda j}\equiv u_{\omega j}$, with $\omega\ge 0$ for $\lambda=\omega^2\ge 0$,  
	and $\omega<0$ for
	$\lambda=-\omega^2<0$. 
	Since the second term of Eq. \eqref{H2} will be infinite summation of terms as $\lambda \phi_\lambda ^2$, the eigenvalue of the Hamiltonian can take any negative value and can not be bounded from below once $\lambda$ can take negative value. Furthermore, for $\lambda<0$, the time dependence of the mode function can be easily seen from the above Eq. \eqref{eom1}, as the form of $e^{at},\:a\in\mathbb{R}$, which  is called  classical instability \cite{TM}. According to \cite{Wald}, for $\xi\le \frac{9}{48}$, the eigenvalue of $A$ can be non-negative and the Hamiltonian can be bounded from below via positive self-adjoint extensions. This will be discussed in details in sec. 2.2. For $\xi > \frac{9}{48}$, no positive self-adjoint extensions and the spectra of $A$ and hence of $H$ have no minimum. However, in the framework of canonical quantization, for any $\xi$, the negative part of the spectrum of the Hamiltonian, as well as the 'classical instability', can be made trivial, leading to a spectrum with minimum.

	\subsection{The Stability of Quantum Vacuum in AdS}

	The real scalar field operator, which is Hermitian, can be expanded as\footnote{The integration of the second term of Eq. \eqref{expansion4} is $\lim_{\epsilon\to 0}\int_{-\infty}^{-\epsilon} d\omega \cdots$, in the following we simply denote as $\int_{-\infty}^{0}$. },
	\begin{equation}\label{expansion4}
		\begin{split}
			\phi(t,r,\theta,\varphi)=&\int_{0}^{+\infty} d\omega \sum_{j,m}N_{\omega}\left[a_{\omega j m}e^{-i\omega t}u_{\omega j}(r)Y_{jm}(\theta,\varphi)+a_{\omega  j m}^\dagger e^{i\omega t} u_{\omega j}(r)Y_{jm}^*(\theta,\varphi)\right]\\
			&+\int_{-\infty}^{0^-}d\omega  \sum_{j,m}N_{\omega}\left[a_{\omega j m}e^{\omega t}u_{\omega j}(r)Y_{jm}(\theta,\varphi)+ a_{\omega j m}^\dagger e^{\omega t}u_{\omega j}(r)Y_{jm}^*(\theta,\varphi)\right.\\
			&\:\:\:\:\:\:\:\:\:\:\:\:\:\:\:\:\:\:\:\:+\left.b_{\omega j m}e^{-\omega t}u_{\omega j}(r)Y_{jm}(\theta,\varphi)+b_{\omega j m}^\dagger e^{-\omega t}u_{\omega j}(r)Y_{jm}^*(\theta,\varphi)\right], \\
		\end{split}
	\end{equation}
	where the normalization constant $N_{\omega}=
	\sqrt{\frac{l}{2\left|\omega\right|}}$. Here we do not specify the concrete value of $\xi$ and boundary condition for $u_{\omega j}$. 
	In some special boundary condition in the summation of $\int_{-\infty}^0d \omega$, some or all (see sec. 2.2) the $u_{\omega j}$ could be zero. \footnote{The summation of $\omega$ might be discrete or continuous depending on the eigenvalue spectrum of $A$ under specific boundary condition. In particular, the above integral of the negative spectrum, can be considered as zero for the case $\xi \le 5/48$, and can only include several finite $\omega<0$ values when $\frac{5}{48}<\xi\le \frac{9}{48}$. }
	
	With this expansion of the field operator,  the Hamiltonian can be expressed by  the  coefficient operators. The part corresponding to the spectrum $\omega \geq 0$ is
	straightforwardly diagonal. On the other hand, for the negative part, we need to introduce the linear transformation
	\begin{equation}\label{2.11z}
		\begin{split}
			&A_{\omega jm}
			=
			\frac{1}{2\sqrt{2}}
			\left[
			a_{\omega jm}+b_{\omega jm}
			+(-1)^m
			\left(
			a_{\omega j,-m}^{\dagger}
			+b_{\omega j,-m}^{\dagger}
			\right)
			\right],\forall m\in[1,j], \\
			&B_{\omega jm}=\frac{1}{2\sqrt{2}}\left[a_{\omega jm}-b_{\omega jm}+(-1)^m\left(a_{\omega j,-m}^{\dagger}-b_{\omega j,-m}^{\dagger}\right)\right],\forall m\in[1,j]\\
			&A_{\omega j0}=\frac{1}{4} \left(a_{\omega j 0}+b_{\omega j 0}+ a_{\omega j 0}^\dagger+b^\dagger_{\omega j 0}\right) ,~B_{\omega j0}
			=
			\frac14
			\left(
			a_{\omega j0}
			+a_{\omega j0}^{\dagger}
			-b_{\omega j0}
			-b_{\omega j0}^{\dagger}
			\right).
		\end{split}
	\end{equation}
	This lead to the relation between $m$ and $-m$
	\begin{equation}\label{m-m}
		A_{\omega j,-m}
		=
		(-1)^m A_{\omega jm}^{\dagger},
		\qquad
		B_{\omega j,-m}
		=
		(-1)^m B_{\omega jm}^{\dagger}. 
	\end{equation}
	Taking into account these relations, the whole Hamiltonian is diagonal
	\begin{equation}\label{instabilityH}
		\begin{split}
			H
			=&\int_0^{+\infty}d\omega\sum_{j, m}\frac{\omega}{2}\{a_{\omega jm},  a_{\omega jm}^{\dagger}\}
			-\int_{-\infty}^{0}d\omega\,
			\omega
			\sum_{j=0}^\infty
			4
			\left(
			B_{\omega j0}^2-A_{\omega j0}^2
			\right)\\
			&-\int_{-\infty}^0d\omega 2\omega\sum_{j=1}^\infty\left[
			{\sum_{m_1}}
			\{B_{\omega jm},B_{\omega jm}^{\dagger}\}
			-\sum_{m_2}
			\{A_{\omega jm},A_{\omega jm}^{\dagger}\}\right], 
		\end{split}
	\end{equation}
	where the range of both of summation index $m_1,m_2$ are uncertain and can be either $[1,j]$ or $[-j,-1]$ because of Eq. \eqref{m-m}. 
	Correspondingly, the second term of the expansion of field Eq. \eqref{expansion4} becomes
	\begin{equation}\label{h}
		\begin{split}
			&\int_{-\infty}^0d\omega N_\omega  \sum_{j=0}^\infty u_{\omega j}\left[{\sum_{m_1}}2\sqrt{2}\left[A_{\omega j m}\cosh \omega tY_{jm}+h.c.\right]+2\sqrt{2}{\sum_{m_2}}\left[B_{\omega j m}\sinh \omega tY_{jm}+h.c.\right]\right.\\
			&\qquad\qquad\qquad\qquad\qquad\left.+4 \left(A_{\omega j 0}\cosh \omega t+B_{\omega j 0}\sinh \omega t\right)Y_{j0}\right], 
		\end{split}
	\end{equation}
	where the summation index $m_1,\:m_2$ again is uncertain as well.

Hence, the negative sector involves independent degrees of freedom $B$ with positive energy and $A$ with negative energy. 
The relations between  the  quantum number $m$ and $-m$ for negative modes, 
$A_{\omega j m}^\dagger=(-1)^mA_{\omega j-m},\:B_{\omega j m}^\dagger=(-1)^mB_{\omega j-m}$, resulting from   the real exponential time dependence which is the 'classical instability',  straightforwardly lead  these  degrees of freedom  as constrained system.  
Physically,    the  direction of the angular momentum of the negative modes is uncertain since either $m\in [0,j]$ or  $m\in [-j,0]$ is opposite in direction. This has been showed in Eq. \eqref{instabilityH} and \eqref{h}. 
As a matter of fact, the relation $A_{\omega j m}^\dagger=(-1)^mA_{\omega j-m},\:B_{\omega j m}^\dagger=(-1)^mB_{\omega j-m}$ makes the  Poisson bracket and commutation relation $[A_m,A_{m'}^\dagger],\:[B_m,B_{m'}^\dagger]$ indefinite. Further, these relations lead to that the non-vanishing and time-dependent equal-time commutation $[\phi(t,r,\theta,\varphi),\phi(t,r',\theta',\varphi')]\ne 0$, which also implies the constraints and the causality and locality violation, if we assign $[A_{\omega j m},A_{\omega' j' m'}^\dagger]=\delta(\omega -\omega')\delta_{jj'}\delta_{mm'},\:[B_{\omega j m},B_{\omega' j' m'}^\dagger]=\delta(\omega -\omega')\delta_{jj'}\delta_{mm'},\:\forall m,m'>0$. Moreover, commutation relations of these coefficient operators always render that the diagonalized Hamiltonian can take any positive or negative eigenvalue and is not bounded from below thus the definition of vacuum is lost.
%

The only other algebraic relation accommodated in 3-dimensional space consistent with locality is the anti-commutation relation, i.e.,  $\forall \omega, \omega ' < 0$\footnote{For $m=0,\omega <0$,  $A_{\omega j 0}=A^\dagger_{\omega j 0}$ and $B_{\omega j 0}=B^\dagger_{\omega j 0}$, namely $A_{\omega<0 j 0}$ and $B_{\omega<0 j 0}$ are Hermitian, and $A_{\omega<0 j 0}^2=0,\:B^2_{\omega<0 j 0}=0$ due to the anti-commutation, so $A_{\omega<0 j 0}=B_{\omega<0 j 0}\equiv 0$. }
\begin{equation}\label{aca}
\begin{split}
	&\left\lbrace A_{\omega j m}, A^\dagger_{\omega^\prime j^\prime m^\prime}\right\rbrace =\delta(\omega-\omega^\prime)\delta_{jj^\prime}\delta_{mm^\prime} ~(m,\:m^\prime\ge 1),\:\mathrm{others\:are\:}0,\\
		&\left\lbrace B_{\omega j m}, B^\dagger_{\omega^\prime j^\prime m^\prime}\right\rbrace =\delta(\omega-\omega^\prime)\delta_{jj^\prime}\delta_{mm^\prime} ~(m,\:m^\prime\ge 1),\:\mathrm{others\:are\:}0, 
\end{split}
\end{equation}
fortunately lead the operator form in the bracket of the second term (negative spectrum) of the Hamiltonian Eq. \eqref{h} trivialized to be a constant, independent from any operators (the constants of $A$ and $B$ in fact cancel)\footnote{Recall that the triviality is one of the well-known basic arguments why commutation relation (bosonic statistics) rather than  anti-commutation relation should be introduced for  any  fields with integer spin in Minkowski spacetime.}. 
Hence all the above problems, constrains, ambiguities, as well as the  vacuum disaster of the negative degrees of freedom are  simultaneously solved by this prescription of anti-commutation relations to the negative spectrum. We would like to address the constrain between $m$ and $-m$ Eq. \eqref{m-m} does not lead to any uncertainty for the anti-commutation relations Eq. \eqref{aca}, contrary to the commutation relations.

The non-negative spectrum is as usual,
\begin{equation}\label{ca}
	\left[ a_{\omega j m}, a^\dagger_{\omega^\prime j^\prime m^\prime}\right ] =\delta(\omega-\omega^\prime)\delta_{jj^\prime}\delta_{mm^\prime},\:\mathrm{others\:are\:}0. 
\end{equation}
All the above equations lead the Hamiltonian to be 
\begin{equation}
	\label{scaham}
	H=\int_0^{+\infty}d\omega\sum_{j, m}\frac{1}{2}\omega \left(a^\dagger_{\omega j m}a_{\omega j m}+a_{\omega j m}a^\dagger_{\omega j m}\right).
\end{equation}
This suggests that after quantization, the spectrum of scalar field with any non-minimal coupling (i.e., $\forall \xi$) in AdS become a bundle of scalar bosons with non-negative  $\omega$ and arbitrary angular momentum quantum number $(j,m)$,  whose Fock state vectors span the Hilbert space in our study. 
Namely,  $a_{\omega j m}^\dagger$ or $a_{\omega j m}$ appears in the Hamiltonian  Eq. \eqref{scaham}
is creation or annihilation operator for the physical boson. While
the negative spectrum   trivialized as above   will never be  active degree of  freedom in this Hilbert space and will be completely  absent
from  the quantum dynamics  by our constructions in the following.  Only because of their 'wrong' algebraic relations   we refer  them as 'ghost particles'. There will be NO need of any 'ghost particle dynamics' (e.g. see the following Eq. \eqref{chi}).

The vacuum state $\left| 0\right\rangle$ is defined as
\begin{equation}\label{v}
	\begin{split}
		a_{\omega j m}\left| 0\right\rangle &=0, ~~\forall \omega \geq 0,j\ge 0\:,-j\le m\le +j. 
	\end{split}
\end{equation}
and the Fock states of the bosons as
\begin{equation}\label{excited}
	\prod_i \frac{\left(a_{\omega_1 j_1 m_1}^\dagger\right)^{n_1}}{\sqrt{n_1!}}\cdots\left| 0\right\rangle, 
\end{equation}
which span the physical Hilbert space.

If there is no restrictions between  the angular quantum numbers with the energy,  and when the minimum physical energy is zero, the vacuum, as the definition correponding to the minimal of Hamiltonian, could be degenerated. The degenerated states contain any number of zero energy physical bosons with arbitrary $j$ and $m$.  This should be the infrared problem as the case of electromagnetic field. 
Such degenerated vacuum can lead to anomaly of the conservation of some quantum number once the detailed balance is destroyed by nonequilibrium. We will not discuss such complexity here and consider the continuous spectrum as a limitation of discrete spectrum.

Now, we further check the consistency by studying the elimination of the infinities in the Hamiltonian
as well as in the energy momentum tensor.
The canonical energy-momentum tensor can be obtained
via the functional derivative of the action with respective to the metric tensor without variation of Ricci scalar curvature\footnote{The energy-momentum tensor taking into account the variation of Ricci scalar curvature will not change the quantization result, the details see Appendix \ref{improvememt}. }, 
\begin{equation}\label{TMUNU}
	T^{\mu\nu}=\frac{1}{2}\left(D^\mu \phi D^\nu \phi+D^\nu \phi D^\mu \phi\right)-\frac{1}{2}g^{\mu\nu}D^\rho\phi D_\rho \phi-\frac{1}{2}g^{\mu\nu}\xi R\phi^2. 
\end{equation}
By contracted with the time-like Killing vector, this energy momentum tensor gives the Hamiltonian Eq. \eqref{H21}, $
	H=\int drd\theta d\varphi \sqrt{-g}\xi_\mu T^{\mu 0}$, where $\xi_\mu=(1+l^{-2}r^2,0,0,0)$ is the timelike Killing vector in AdS.

As usual, divergences should be subtracted via a proper renormalization scheme with the reasonable renormalization condition. Since the vacuum state corresponds to the AdS background, which is the solution of the homogenous Einstein equations, hence the vacuums satisfy the renormalization condition
\begin{equation}\label{RP}
	\left\langle0|T^{\mu\nu}_R|0\right\rangle =0, 
\end{equation}
from which the counterterms can be extracted. Once the $\Lambda_0$ is renormalized to $\Lambda$ in Eq. \eqref{eeq}, the consistency of the vacuum is obtained. With the expansion of the scalar field Eq. \eqref{expansion4} and the (anti-)comutation relations, the vacuum expectation of the energy momentum tensor operator is calculated. The details of the calculation is described in Appendix \ref{appendix}. 

The counterterm is
\begin{equation}\label{CT}
	\begin{split}
		&\Delta T^{\mu\nu}=\Delta T^{\mu\nu}_1+\Delta T^{\mu\nu}_2+\Delta T^{\mu\nu}_3\\
		&\Delta T^{\mu\nu}_1=-\int_0^\infty d\omega\sum_{j,m}\frac{l}{4\omega}(g^{\nu\rho}g^{\mu\sigma}+g^{\mu\rho}g^{\nu\sigma})\left(\partial_\rho\phi_{\omega j m}\partial_\sigma \phi_{\omega j m}^*\right)\\
		&\Delta T^{\mu\nu}_2=g^{\mu\nu}\left[\int_0^\infty d\omega\sum_{j,m}\frac{l}{4\omega}\left(g^{\rho\sigma}\partial_\rho \phi^*_{\omega j m}\partial_\sigma\phi^*_{\omega j m}+\xi R\phi_{\omega j m}^*\phi_{\omega j m}\right)\right]\\
		&\Delta T^{\mu\nu}_3=\sum_{i=1,2}\int_{-\infty}^0\frac{d\omega l}{2\omega}\sum_{j=1}^{+\infty}\sum_{m=1}^jf^{\mu\nu(i\circ)(i*)}_{\omega j m\omega j m}. 
	\end{split}
\end{equation}

The renormalization by introducing $\Delta T_1^{\mu\nu}$ and $\Delta T_2^{\mu\nu}$ is equivalent to taking normal ordering to the bosonic sector of $T^{\mu\nu}$. 
The last term $\Delta T^{\mu\nu}_3$ ensures the cancellation of the unphysical contributions from the ghost operators, 
so that the renormalized energy-momentum tensor $T^{\mu\nu}_R=T^{\mu\nu}+\Delta T^{\mu\nu}$ is independent of the ghost. 
Based on the relation of the unrenormalized Hamiltonian and the energy momentum tensor pointed above, the renormalized Hamiltonian can be obtained via
\begin{equation}\label{HT}
	H_R=\int drd\theta d\varphi \sqrt{-g}\xi_\mu T^{\mu 0}_{R}. 
\end{equation}
Hence we have 
\begin{equation}
	\begin{split}
		H_R&=\int_0^{+\infty} d\omega \sum_{j=0}^{+\infty}\sum_{m=-j}^j\frac{1}{2}\omega \left(a^\dagger_{\omega j m}a_{\omega j m}+a_{\omega j m}a^\dagger_{\omega j m}\right)+\int drd\theta d\varphi r^2\sin\theta\left(1+l^{-2}r^2\right)(\Delta T^{\mu\nu}). 
	\end{split}
\end{equation}
Here the contribution of $\Delta T_1^{\mu\nu}+\Delta T_2^{\mu\nu}$ cancels the divergence in non-negative spectrum, and the contribution of $\Delta T_3^{\mu\nu}$ vanishes, which leads to the
renormalized Hamiltonian, 
\begin{equation}
	H_R=\int_0^{+\infty} d\omega \sum_{j=0}^{+\infty}\sum_{m=-j}^j\omega a^\dagger_{\omega j m}a_{\omega j m}. 
\end{equation}
We can see that the renormalization condition, Eq. \eqref{RP}, has compeletly determined the renormalization of the Hamiltonian. $H_R$ is bounded from below so the vacuum is stable. The expectation values of both $H_R$ and $T_{R}^{\mu\nu}$ on excited states Eq. \eqref{excited} are finite. 
The Hamiltonian is different from the energy momentum tensor in that it is a integrated quantity (The surface term Eq. \eqref{ST} will be investigated in details with concrete boundary conditions in next subsection. ).

It is easy to show that any expectation values of $T^{\mu\nu}_R$ on any Fock state is covariantly conserved
\begin{equation}
	D_\mu \left\langle T^{\mu\nu}_R\right\rangle :=\partial_\mu \left\langle T^{\mu\nu}_R\right\rangle +\Gamma_{\mu \rho}^\mu \left\langle T^{\rho\nu}_R\right\rangle +\Gamma_{\mu \rho}^{\nu}\left\langle T^{\mu\rho}_R\right\rangle=0. 
\end{equation}
This is consistent with Einsetin equation
\begin{equation}\label{nablatmunu}
	D_\mu \left(G^{\mu\nu}+\Lambda g^{\mu\nu}\right)=D_\mu \left\langle T^{\mu\nu}_R\right\rangle =0. 
\end{equation}

The above investigation show the ghost degree of freedom are not active in the dynamics of energy-monetum tensor either.  With the well-defined Hilbert space completely spanned by physical particle Fock states in hand, we now construct regular quantum fields, completely free from ghost, following Weinberg \cite{weinberg1995quantum}. The key point is that, the function space spanned by mode functions corresponding to $\omega\ge 0$ is isomorphic to the Hilbert space spanned by Fock states Eq. \eqref{excited}. 
	
The physical Fock states are the unitary representation of the AdS symmetry group SO(2,3)
	\begin{equation}
		\begin{split}
			&U_0(\Lambda)a_{\omega j m}U_0^{-1}(\Lambda)=\int_0^\infty d\bar{\omega}\sum_{j=0}^\infty\sum_{m=-j}^j D_{\omega j m;\bar{\omega} \bar{j} \bar{m}}a_{\bar{\omega}\bar{j} \bar{m}},\\
			&U_0(\Lambda)a_{\omega j m}^\dagger U_0^{-1}(\Lambda)=\int_0^\infty d\bar{\omega}\sum_{j=0}^\infty\sum_{m=-j}^j D_{\omega j m;\bar{\omega} \bar{j} \bar{m}}^*a^\dagger_{\bar{\omega} \bar{j} \bar{m}},
		\end{split} 
	\end{equation}
	where $\Lambda\in $ SO(2,3), and $\omega\ge 0$. Now we construct the covariant scalar field, 
	\begin{equation}\label{chi}
		\chi(x)=\int_0^{+\infty} d\omega \sum_{j=0}^\infty \sum_{m=-j}^jN_{\omega}f_\omega\left[a_{\omega j m}e^{-i\omega t}u_{\omega j}(r)Y_{jm}(\theta,\varphi)+a_{\omega  j m}^\dagger e^{i\omega t} u_{\omega j}(r)Y_{jm}^*(\theta,\varphi)\right], 
	\end{equation}
	to guarantee $U_0(\Lambda) \chi(x)U_0^{-1}(\Lambda)=\chi(\Lambda x)$, 
	and $f_\omega$ satisfies
	\begin{equation}
		f_\omega \phi_{\omega j m}(\Lambda x)=\int_0^\infty d\bar{\omega}\sum_{j=0}^\infty\sum_{m=-j}^j D_{\bar{\omega} \bar{j} \bar{m};\omega j m}f_{\bar{\omega}}\phi_{\bar{\omega} \bar{j} \bar{m}}\frac{N_{\bar{\omega}}}{N_\omega}. 
	\end{equation}
	Following the isomorphism between Hilbert space and mode function space, we have 
	\begin{equation}\label{wb}
		\int_0^\infty d\omega \sum_{j,m}f_\omega^2u_{\omega j}(r)Y_{jm}(\theta,\varphi)u_{\omega j}(r')Y^*_{jm}(\theta',\varphi')=\frac{\delta(r-r')\delta(\theta-\theta')\delta(\varphi-\varphi')(1+l^{-2}r^2)}{lr^2\sin\theta}. 
	\end{equation}
	
	From Eq. \eqref{chi}, the equation of motion of $\chi$ has the similar form as Eq. \eqref{F} $\hat{F}[\chi]=0$, so the Lagrangian of $\chi$ similar to \eqref{lagrangian} with $\phi$ replaced by $\chi$, resulting in the canonical equal-time commutation 
	\begin{equation}
		[\chi(t,r,\theta,\varphi),\Pi_\chi(t,r',\theta',\varphi')]=i\delta(r-r')\delta(\theta-\theta')\delta(\varphi-\varphi'),\:\Pi_\chi=\frac{r^2\sin \theta}{1+l^{-2}r^2}\dot{\chi}. 
		\end{equation}
	With this Lagrangian and Eq. \eqref{wb}, we can obtain the Hamiltonian of $\chi$ is similar to Eq. \eqref{H2}
	\begin{equation}
		:H_\chi:=\frac{1}{2}\int drd\theta d\varphi \frac{r^2\sin\theta}{1+l^{-2}r^2}:\left(\dot{\chi}^2-\ddot{\chi}\chi\right):=\int_{0}^\infty d\omega\sum_{j=0}^\infty\sum_{m=-j}^j\omega a^\dagger_{\omega j m}a_{\omega j m}. 
	\end{equation}
	Hence, the Heisenberg equation is
	\begin{equation}
		\dot{\chi}=\frac{1}{i}[\chi,H_\chi],\:\dot{\Pi}_\chi=\frac{1}{i}[\Pi_\chi,H_\chi], 
	\end{equation}
	which guarantee unitary evolution. 
	The canonical energy-momentum tensor has the similar form Eq. \eqref{TMUNU} with $\phi$ replaced by  $\chi$, so $\nabla_\mu \left\langle :T_{\chi}^{\mu\nu}:\right\rangle =0$. 
	From the above construction, the free quantum field theory of $\chi$ satisfies causality and locality. These properties lead the Feynman propagator $i\Delta_F(x,y)=\left\langle 0\right| T[\chi(x),\chi(y)] \left|0 \right\rangle$ satisfies 
	\begin{equation}\label{feynman}
		\hat{F}\left\langle 0\right| T[\chi(x),\chi(y)] \left|0 \right\rangle =-\frac{1}{\sqrt{-g}}\delta(x^0-y^0)[\chi(t,y^i),\Pi_\chi(t,x^i)]= -i\delta^4(x,y), 
	\end{equation}
	here $H_\chi \left| 0\right\rangle =0$. It is clear that the isomorphism between Hilbert space and the mode function space plays an important r\^ole. 
	
	With this self-consistent field theory in hand, 
	as pointed by Weinberg \cite{weinberg1995quantum}, various covariant interaction terms can be introduced to construct EQFT. For a specfic EQFT with the Lagrangian $\mathcal{L}(\tilde{\phi})+\mathcal{L}_I(\tilde{\phi})$, the equation of motion becomes
	\begin{equation}
		\hat{F}[\tilde{\phi}]=-J,\:J(x)=\frac{\partial \mathcal{L}_I}{\partial \tilde{\phi}(x)}. 
	\end{equation}
	
	The relation between the interaction field $\tilde{\phi}$ and free field $\chi$ can be described by Yang-Feldman-like equation\footnote{As is well known, in practical calculation, the equation should be used in matrix element form rather than the original operator form \cite{Yang:1950vi}. } \cite{Yang:1950vi, Schroer:2004yr, Gottschalk:2006pj, Woodard:2006pw, Onemli:2015pma}
	\begin{equation}\label{yf}
		\tilde{\phi}(x)=\chi(x)-\int d^4 x' D(x,x')J(x'),
	\end{equation}
	where $D$ is a Green function satisfying $\hat{F}[D(x,x')]=\delta^4(x,x')$ and determined by a specific boundary condition. For example, if the $D$ is the retarded Green function, the $\chi$ can be considered as the in-field of the EQFT with Lagrangian $\mathcal{L}(\tilde{\phi})+\mathcal{L}_I(\tilde{\phi})$. Since the operator $\hat{F}$ covers the full spectrum $(-\infty,+\infty)$ for $\xi>5/48$, in Appendix \ref{c}, we illustrate how to keep the Green function well defined in the mode function space with $\omega \ge 0$, which is crucial to keep the interacting field $\tilde{\phi}$ belong to the mode function space, completely free from the ghost degrees of freedom.

\subsection{Quantization with Positive Self-adjoint Extension}
According to Wald \cite{Wald}, for three cases corresponding to the range of $\xi$, namely $\xi\le 5/48,\:5/48<\xi<9/48,\:\xi=9/48$\footnote{
	The massive cases could be obtained by the similar method just substituting $\xi$ with $\xi-\frac{m^2l^2}{12}$. Similar conclusions could be obtained by replacing  $\xi\le \frac{5}{48},\:\frac{5}{48}<\xi	< \frac{9}{48}	,\xi=\frac{9}{48}$ by $\xi\le \frac{5}{48}+\frac{m^2l^2}{12},\:\frac{5}{48}+\frac{m^2l^2}{12}<\xi< \frac{9}{48}+\frac{m^2l^2}{12},\xi=\frac{9}{48}+\frac{m^2l^2}{12}$ respectively. Here we do not discuss the details of the meaning of $m^2$ which is a free parameter taking real value.}, 
the positive self-adjoint extensions of $A$ exist. There is uniquely positive self-adjoint extensions for $\xi\le \frac{5}{48}$ and in fact in this case always $\lambda>0$. 
There exists a family of parameters (boundary conditions corresponding to radial direction) that make $A$ (and Hamiltonian) is bounded from below in $\frac{5}{48}<\xi<\frac{9}{48}$ and $\xi=\frac{9}{48}$, respectively. 

\subsubsection{$\xi\le \frac{5}{48}$}
The operator $A$ of Eq. \eqref{A} has discrete eigenvalues
\begin{equation}\label{lambda}
	\lambda_{nj}=l^{-2}\left(2n+j+\frac{3}{2}+\sqrt{\frac{9}{4}-12\xi}\right)^2,\:n=0,1,2,\cdots
\end{equation}
which guarantees that the $u_{n j}$'s are square integrable. \footnote{See \cite{Wald} for further detials about the explicit solutions. The relation between $u_{nj}(r)$ here and $\Phi_{\omega j}(x)$ in Wald's paper \cite{Wald} is $u_{nj}=l^{-1/2}\frac{\Phi_{\omega j}}{r}$. }.

The normalization condition of $u_{\omega j}$ can be written as,
\begin{equation}
	\int_0^\infty u_{n j}u_{n^\prime j}\frac{r^2dr}{1+l^{-2}r^2}=\delta_{nn^\prime}.\:
\end{equation}
The real scalar field is expanded by the orthonormal completeness set of the mode functions,
\begin{equation}\label{expansion}
	\begin{split}
		\phi(t,r,\theta,\varphi)=\sum_{n ,j ,m}&\frac{1}{\sqrt{2}\lambda_{nj}^{1/4}}\left[ a_{n j m}e^{-i\sqrt{\lambda}t}u_{n j}(r)Y_{jm}(\theta,\varphi)+a_{n j m}^\dagger e^{i\sqrt{\lambda}t} u_{n j}(r)Y_{jm}^*(\theta,\varphi)\right]. 
	\end{split}
\end{equation}
The Hamiltonian Eq. \eqref{H2} is
\begin{equation}
	H=\sum_{n j m}\frac{\sqrt{\lambda}}{2}\left(a_{n j m}a^\dagger_{n j m}+a^\dagger_{n j m}a_{n j m}\right). 
\end{equation}
The term Eq. \eqref{ST} are explicitly calculated to be $0$. 
We introduce the canonical commutation relations consistent with the diagonal Hamiltonian, 
\begin{equation}
	[a_{n j m},a^\dagger_{n^\prime j^\prime m^\prime}]=\delta_{n,n^\prime}\delta_{jj^\prime}\delta_{mm^\prime},\:others \:are\:zero, 
\end{equation}
i.e.,  $a_{n j m}$ or $a^\dag_{n j m}$ is the canonical annihilation or creation operator, respectively.
This means the positive but not too large $\xi$ ($\le \frac{5}{48}$)
values do not change the non-negative property of the Hamiltonian though the term $\xi R\phi^2$ introducing negative contributions. 

The $\phi$ and $\Pi$ are canonical variables\footnote{We could obtain the familiar result $[\phi(\mathbf{x},t),\:\Pi(\mathbf{x}^\prime,t)]=i\delta^3(\mathbf{x-x}^\prime)$ if the Cartesian coordinate is employed.},
\begin{equation}
	\left[\phi(r,\theta,\varphi, t),\Pi(r^\prime, \theta^\prime,\varphi^\prime, t)\right]=i\delta(r-r^\prime)\delta(\theta-\theta^\prime)\delta(\varphi-\varphi^\prime). 
\end{equation}
Hence the standard  Heisenberg equations lead to the quantum equation of motion exactly
the similar form as the classical one. The quantum effective Lagrangian density share the similar form as the classical one.

As usual, the unique vacuum could be defined by $a_{n j m}\left| 0\right\rangle =0,\:\forall n j m$. The particle states can be unambiguously defined  by the creation operators acting on vacuum state and it is the eigenstate of the Hamiltonian, which construct the Fock space. $\left\langle \Psi|H|\Psi\right\rangle \ge 0$  with state vector $\left|\Psi\right\rangle$  in the Fock space and the equal sign for and only for the quantum vacuum state.
Hence the quantum vacuum is stable and the quantum effective  dynamics of the scalar  fields   described with the Lagrangian  (Eq. \eqref{lagrangian}) with $\xi\le 5/48$
is well defined in the AdS spacetime.

The similar renormalization scheme with the renormalized condition as Eq. \eqref{RP} in sec. 2.1 can be performed. The counter term can be calculated to be
\begin{equation}\label{counterterms1}
	\begin{split}
		&\Delta T^{\mu\nu}=\Delta T^{\mu\nu}_1+\Delta T^{\mu\nu}_2 \\
		&\Delta T^{\mu\nu}_1=-\sum_{n,j,m}\frac{1}{4\sqrt{\lambda}}(g^{\nu\rho}g^{\mu\sigma}+g^{\mu\rho}g^{\nu\sigma})\left(\partial_\rho\phi_{\omega j m}\partial_\sigma \phi_{\omega j m}^*\right)\\
		&\Delta T^{\mu\nu}_2=g^{\mu\nu}\left[\sum_{n,j,m}\frac{1}{2}\left(g^{\rho\sigma}\partial_\rho \phi^*_{n j m}\partial_\sigma\phi^*_{n j m}+\xi R\phi_{n j m}^*\phi_{n j m}\right)\right], 
	\end{split}
\end{equation}
and the renormalized energy-momentum tensor is
\begin{equation}\label{RT1}
	T_{R}^{\mu\nu}=T^{\mu\nu}+\Delta T^{\mu\nu}=:T^{\mu\nu}:, 
\end{equation}
i.e., the divergences in the energy momentum tensor can be completely removed by taking the normal order of the operators, so does the renormalized Hamtonian
\begin{equation}\label{H5/48}
	H_R=\sum_{n j m}l^{-1}\left(2n+j+\frac{3}{2}+\sqrt{\frac{9}{4}-12\xi}\right)a^\dagger_{n j m}a_{n j m}=:H:. 
\end{equation}
From the above, the free field theory in AdS can be made finite by taking the normal order of the operators in AdS in case of ghost(negative spectrum)-free, which is similar as in Minkowski spacetime. And the concept of particle is well defined in the whole AdS.

The energy spectrum is discrete for $\xi\le \frac{5}{48}$. There is an energy-gap between the vacuum and the  one particle states with the smallest energy ($n=0,\:j=0,\:\forall m$). The backreaction perturbations initialized by the various particle states are not infinitesimal, which is different from the classical consideration. 
\subsubsection{$\:\frac{5}{48}< \xi < \frac{9}{48}$ and $\xi=9/48$}
Adopting the positive self-adjoint extensions of the operator $A$ Eq. \eqref{A}, corresponding to certain boundary conditions, the $H$ has non-negative spectrum. Here we denote continuous  $\lambda=\omega^2,\:\omega\ge 0$.
Similar quantization and renormalization scheme as the cases in $\xi\le \frac{5}{48}$ can be performed. The counterterms of energy-momentum tensor are
\begin{equation}\label{counterterms2}
	\begin{split}
		&\Delta T^{\mu\nu}=\Delta T^{\mu\nu}_1+\Delta T^{\mu\nu}_2\\
		&\Delta T^{\mu\nu}_1=-\int_0^\infty d\omega\sum_{j,m}\frac{l}{4\omega}(g^{\nu\rho}g^{\mu\sigma}+g^{\mu\rho}g^{\nu\sigma})\left(\partial_\rho\phi_{\omega j m}\partial_\sigma \phi_{\omega j m}^*\right)\\
		&\Delta T^{\mu\nu}_2=+g^{\mu\nu}\left[\int_0^\infty d\omega\sum_{j,m}\frac{l}{4\omega}\left(g^{\rho\sigma}\partial_\rho \phi^*_{\omega j m}\partial_\sigma\phi^*_{\omega j m}+\xi R\phi_{n j m}^*\phi_{n j m}\right)\right]. 
	\end{split}
\end{equation}

The renormalized energy-momentum tensor and Hamiltonian is same as Eq. \eqref{RT1} and Eq. \eqref{H5/48} respectively, i.e., can be obtained by taking normal ordering of the operator products. 

The boundary term Eq. \eqref{ST} is 
\begin{equation}
	\begin{split}
		R^2(1+l^{-2}R^2)\int \frac{ld\omega d\omega^\prime}{4\sqrt{\omega\omega^\prime}}\sum_{jm}&\left[a_{\omega jm}a_{\omega^\prime j-m} e^{-i(\omega+\omega^\prime)t}u_{\omega j}(R)u_{\omega^\prime j}^\prime(R)(-1)^m\right. \\
		&+a_{\omega jm}^\dagger a_{\omega^\prime j-m}^\dagger e^{i(\omega+\omega^\prime)t}u_{\omega j}(R)u_{\omega^\prime j}^\prime(R)(-1)^m\\
		&+a_{\omega jm}^\dagger a_{\omega^\prime jm} e^{i(\omega-\omega^\prime)t}u_{\omega^\prime j}(R)u_{\omega j}^\prime(R)\\
		&\left.+a_{\omega jm}^\dagger a_{\omega^\prime jm} e^{i(\omega-\omega^\prime)t}u_{\omega^\prime j}(R)u_{\omega j}^\prime(R)\right], 
	\end{split}
\end{equation}
where $R=+\infty$, and $u_{\omega j m}(R)$ and $u^\prime_{\omega j m}(R)$ get value from the concrete radial boundary conditions of mode functions $u_{\omega j m}$. 
The expectation of vacuum $\left|0\right\rangle$ of $T_{R}^{\mu\nu}$ and $H_R$, with the latter including the term Eq. \eqref{ST}, still be $0$, under the renormalization scheme here. This implies that the vacuum of real scalar field under the boundary conditions for the positive self-adjoint extensions can be well defined in whole AdS, and that the Einstein equation in this vacuum state is homogeneous $G^{\mu\nu}+\Lambda g^{\mu\nu}=\left\langle 0|T_R^{\mu\nu}|0\right\rangle =0$. For the excited states, the expectation values of $T_R^{\mu\nu}$ are finite and covariantly conserved; those of the Hamiltonian is the summation of Eq. \eqref{H2} and Eq. \eqref{ST}. The extra contribution on the radial boundary $r=+\infty $ to the Hamiltonian is 
\begin{align}\label{STC}
	\text{the extra energy}\sim
	\begin{Bmatrix}
		O(R^{2\left(\sqrt{\frac{9}{4}-12\xi}\right)}),\:\frac{5}{48}<\xi<\frac{9}{48}\\\frac{1}{R}\frac{R^2-l^2}{\sqrt{R^2+l^2}}\frac{-\ln \left(1+l^{-2}R^2\right)}{-\frac{b_0}{a_0}\ln(1+l^{-2}R^2)+1},\:\xi=\frac{9}{48}
	\end{Bmatrix}, 
\end{align}
where 
$a_0$ and $b_0$ and the concrete formula can be obtained by the concrete radial boundary condition \cite{Wald}. Hence the Hamiltonian can be split into two parts, convergent bulk term \eqref{H2} and boundary term \eqref{ST} that might be divergent depending on the boundary conditions. If it exists the boundary condition leading to the extra energy vanishing, then the energy of particle just is $\omega$. However, the particle energy is a subtle problem when boundary term is divergent.

\section{Maxwell (Abelian gauge) field}

The Lagrangian density of the Maxwell field  in AdS Eq. \eqref{MF}  is only dependent on  the 2-form of the field vector $A_\mu$, i.e., the gauge invariance under transformation $A_\mu \to A_\mu+\partial_\mu \phi$ is still respected.  So similar  as in Minkowski spacetime, redundant degrees of freedom
leads to  disasters in  na\"ive  Legendre transformation,  
i.e., the conjugate momentum of the degree of freedom $A_0$ is just 0, since  $\partial_t A_0$ does not appear in the Lagrangian density. At the same time, via na\"ive Legendre transformation
\begin{equation}\label{H}
	\begin{split}
		&\int drd\theta d\varphi \sqrt{-g}\left(\frac{\partial \mathcal{L}}{\partial \dot{A}_i}\dot{A}_i-\mathcal{L}\right)\\&=\int drd\theta d\varphi\sqrt{-g}\left(\frac{1}{2}g^{ij}g^{00}\partial_iA_0\partial_jA_0-\frac{1}{2}g^{ij}g^{00}\partial_0A_i\partial_0A_j+\frac{1}{4}F_{ij}F_{i^\prime j^\prime}g^{ii^\prime}g^{jj^\prime}\right),  
	\end{split}
\end{equation}
we get a `Hamiltonian`, which is neither gauge invariant nor positive definite (hence losing the definition of the vacuum). However, by fixing a gauge these problems are  easy to be solved. Adopting  the temporal gauge, $A_0=0$, the independent degrees of freedom are $A_i$, and the correct Hamiltonian can be obtained,
which equals to the one obtained from the gauge-invariant canonical   momentum-energy tensor
\begin{equation}\label{tmunu}
	T^{\mu\nu}=F_\lambda^{\:\:\:\mu}F^{\lambda \nu}-\frac{1}{4}g^{\mu\nu}F_{\rho\sigma}F^{\rho\sigma}, 
\end{equation}
with the help of  of the timelike Killing vector:
\begin{equation}\label{hammax}
	H=\int drd\theta d\varphi \sqrt{-g}\xi_\mu T^{\mu 0}=\int drd\theta d\varphi \sqrt{-g}\left[-\frac{1}{2}g^{00}g^{ij}F_{i0}F_{j0}+\frac{1}{4}g^{ii^\prime}g^{jj^\prime}F_{ij}F_{i^\prime j^\prime}\right]. 
\end{equation}
It is gauge invariant and non-negative.  In other words,   the canonical quantization can be simply  based on the
Hamiltonian $H$ in Eq. \eqref{hammax}, with the practical way to employ the field vector $A_\mu$ with the temporal gauge.
It is easy to check that, in the AdS, the first order differential equation $\partial_t \phi=-A_0$ has regular solution.


To single out the proper dynamical degree of freedom in AdS, we adopt the expansion of the Maxwell field as \cite{Ruffini, Wh, CV}, and taking the $A_0=0$ gauge,  
\begin{equation}\label{expansion2}
	A_{\mu}=\sum_{j,m}
	\begin{pmatrix}
		0\\
		0\\
		\frac{a^{jm}(r,t)}{\sin\theta}\frac{\partial Y_{jm}}{\partial \varphi}\\
		-a^{jm}(r,t)\sin\theta\frac{\partial Y_{jm}}{\partial\theta}\\
	\end{pmatrix}+
	\begin{pmatrix}
		0\\
		h^{jm}(r,t)Y_{jm}\\
		k^{jm}(r,t)\frac{\partial Y_{jm}}{\partial \theta}\\
		k^{jm}(r,t)\frac{\partial Y_{jm}}{\partial \varphi}
	\end{pmatrix},
\end{equation}
where the first column being of parity $(-1)^j$ and the second $(-1)^{j+1}$.
%

The Hamiltonian leads to the equations
\begin{align}
	\label{eom2-1}&(g^{rr}a^{jm\:\prime})^\prime+g^{00}\frac{\partial^2 a^{jm}}{\partial t^2}-\frac{j(j+1)}{r^2}a^{jm}=0\\
	\label{eom2-2}&g^{rr}b^{jm\:\prime}=\partial_t k^{jm}\\
	\label{eom2-3}&g^{00}\partial_tb^{jm}=h^{jm}-k^{jm\:\prime}
\end{align}
where
\begin{equation}\label{eomb}
	\frac{j(j+1)b^{jm}}{r^2}:=\partial_t h^{jm}.
\end{equation}
The equation of $b^{jm}$ is the same as $a^{jm}$, which can be obtained by combing Eq. \eqref{eom2-2} and Eq. \eqref{eom2-3}.

By separating varibles,  $a^{\lambda jm}(r,t)=a^{\lambda jm}(r)a^{\lambda jm}(t),\:b^{\lambda jm}(r,t)=b^{\lambda jm}(r)b^{\lambda jm}(t)$, 
\begin{align}
	&\frac{d^2a^{\lambda j m}(t)}{dt^2}=-\lambda a^{\lambda j m}(t)\\
	&(g^{rr}a^{\lambda j m \prime}(r))^\prime-\lambda g^{00}a^{\lambda j m}(r)-\frac{j(j+1)a^{\omega j m}(r)}{r^2}=0.
\end{align}
The same as  for $b^{\lambda j m}(r),\:b^{\lambda j m}(t)$.  
Both the time-dependence of $b^{\lambda j m}(t)$ and $a^{\lambda j m}(t)$ are $e^{\pm i\sqrt{\lambda}t}$.
$\lambda$ is non-negetive since the Hamiltonain is positive definite here. $b^{\lambda j m}(r)$ and $a^{\lambda j m}(r)$ satisfy the hyperspherical differential equation which have been studied (see \cite{ito1993encyclopedic}). Utilizing the results above, we can obtain $h^{\omega jm}(r,t)$ and $k^{\lambda j m}(r,t)$ respectively,
\begin{equation}
	\begin{split}
		&h^{\lambda j m}(r,t)=\frac{j(j+1)b^{\lambda j m}(r,t)}{-i\sqrt{\lambda} r^2}+g^{\lambda j m \prime}(r)\\
		&k^{\lambda j m}(r,t)=\frac{(1+l^{-2}r^2)b^{\lambda j m\:\prime}(r,t)}{-i\sqrt{\lambda}}+g^{\lambda j m}(r).
	\end{split}
\end{equation}
$g^{\lambda j m}(r)$ represents the background degrees of freedom and satisfies the equation,
\begin{equation}
	j(j+1)g^{\lambda j m}(r)=g^{rr}(r^2g^{\lambda j m \prime}(r))^\prime, 
\end{equation}
which does not propagate and is not quantized. As a matter of fact, the contribution of $g^{\lambda j m}$ to $A_\mu$ is 
\begin{equation}
	A_\mu^{background}(r,\theta,\varphi)=\int d\lambda \sum_{jm}c_{\lambda j m}\partial_\mu (g^{\lambda j m}Y_{j m})+h.c.. 
\end{equation}
Hence, it does not contribute to $F_{\mu\nu}$ neither to the Hamiltonian. Hence, the Hamiltonian can be obtained from Eq. \eqref{hammax},
\begin{equation}\label{hmaxwell}
	H=\int_0^\infty d\lambda\sum_{j m,i=a,b}\left|N_\lambda\right|^2\lambda j(j+1)\left(c_i^{\lambda j m}c_i^{\lambda j m\dagger}+c_i^{\lambda j m\dagger}c_i^{\lambda j m}\right), 
\end{equation}
where $N_\lambda$ is normalization factor to guarantee the creation and annihilation operators satisfy the canonical form. The dynamical degrees of freedom of Maxwell field in AdS$_4$ are $a^{\lambda j m}, b^{\lambda j m}$ with the parity $(-1)^j$ and $(-1)^{j+1}$ respectively. 

The renormalization condition is $\left\langle 0|T^{\mu\nu}_R|0\right\rangle =0$, which is similar as the scalar case, Eq. \eqref{RP}, and the counterterm is,
\begin{equation}\label{trmunu}
	\begin{split}
		\Delta T^{\mu\nu}=-\left\langle 0|T^{\mu\nu}|0\right\rangle . 
	\end{split}
\end{equation}
$H$ is positive definite, hence ghost sector is not needed, so the counterterm is equivalent to taking normal ordering. 

The renormalized Hamiltonian is obtained from Eq. \eqref{trmunu} with the help of Killing vector,
\begin{equation}
	H_R=\int_0^\infty d\lambda\sum_{j m,i=a,b}2\left|N_\lambda\right|^2\lambda j(j+1)c_i^{\lambda j m\dagger}c_i^{\lambda j m}=:H:, 
\end{equation}
which eliminates the infinity vacuum energy in Eq. \eqref{hmaxwell}.

	Now it is feasible to consider the interaction between the scalar field and Maxwell field by introducing interaction Lagrangian $\mathcal{L}_I[\tilde{\phi}, V]$ and the interacting field $V^\mu$ . The equations of motion are
	\begin{equation}
		\begin{split}
			\hat{F}[\tilde{\phi}]
			&=-\frac{\partial\mathcal{L}_I[\tilde{\phi},V]}{\partial \tilde{\phi}},
			\qquad
			\hat{G} [V^\mu]
			=-\hat{P}^{\mu\rho}\left[\frac{\partial \mathcal{L}_I[\tilde{\phi},V]}{\partial V^\rho}\right]=-J^\mu_{\text{T}}, 
		\end{split}
	\end{equation}
	where $\hat{P}$ is the gauge fixing operator of temporal gauge, the Yang-Feldman equations are
	\begin{equation}
		\begin{split}
			\tilde{\phi}(x)
			&=\chi(x)-\int d^4x'\,
			D(x,x')
			\frac{\partial \mathcal{L}_I[\tilde{\phi},V]}
			{\partial \tilde{\phi}(x')}, \\
			V_T^\mu(x)
			&=A_T^\mu(x)-\int d^4x'\,
			D_V^{\mu\nu}(x,x')\frac{\partial \mathcal{L}_I[\tilde{\phi},V]}{\partial V^\nu}, 
		\end{split}
	\end{equation}
	where the Green function $D_V^{\mu\nu}(x,x')$ satisfies $\hat{G}[{D_V}^{\nu\rho}(x,x')]=\hat{P}^{\nu \rho}\delta^4(x,x')$.

\section{Conclusions and Discussions}

The AdS spacetime is the maximally symmetric solution of the vacuum Einstein equations with a negative cosmological constant.
In this paper, we have studied the canonical quantization for non-minimally coupled real scalar field and Maxwell fields in (covering) AdS spacetime with  the   parameterization Eq. \eqref{lemetric}.
These two quantum dynamics both have the stable vacuum state
consistent with the background AdS geometry,
and the global Hilbert space
defined by the physical particle Fock states.
To get these results, the following are crucial:

The orthonormal mode functions can be used to expand the fields and the Hamiltonian
is diagonalized for reading out the canonical variables and introducing the particle concept.  
Particularly, for the case of $\xi>5/48$, the Hamiltonian can have negative spectrum unbounded from below. 
However, the negative part of the Hamiltonian spectrum is found as constrained systyem. These constrains are intrinsically lies in the real time-dependent evolution of each negative mode, which result in that the direction of angular-momentum and the commutation relation indefinite. The only possible and definite algebraic anti-commutation relation is employed, which trivializes the negative energy spectrum and obtains the well-defined vacuum. As usual, for the non-negative part, we have the physical Hilbert space spanned by these Fock states applying the commutation relation. After the above quantization, the energy-momentum tensor is renormalized to remove the divergences via the  BPHZ-like way through employing the renormalization  condition  $\left\langle 0|T^{\mu\nu}_R|0\right\rangle =0$.
From this finite $T^{\mu\nu}_R$, contracted with the time-like Killing vector, one gets the renormalized  Hamiltonian bounded from below, and the global stable vacuum and stable excited states, hence complete physical Hilbert space are obtained. The above analysis shows that, for non-minimal coupled scalar fields with any $\xi$, the dynamics itself inevitably lead to the constraints and the associated physical Hilbert space. As a matter of fact, the scalar field and vector field in this paper demonstrate two ways of canonical quantization for constrained system. 

It is obvious that the associated physical Hilbert space is isomorphic to the associated function space spanned by the corresponding mode functions. Based on this isomorphism, we construct the new covariant free scalar fields $\chi$ free from the negative degrees of freedom and interaction fields $\tilde{\phi}$ and $V^\mu$ following the framework of Weinberg \cite{weinberg1995quantum}, which is a deep insight of the relation between particle concept and quantized field in Minkowski spacetime. Hence our work can be considered as an extension of this insight to AdS spacetime. The relations between the free fields $\chi$ and $A_\mu$ and the interaction scalar field $\tilde{\phi}$ and interacting Maxwell fields $V^\mu$ are described by the well-known Yang-Feldman equations respectively. 

The quantization procedure of our paper aligns with the present works such as \cite{AS, BF, BF2, DD, BM, BJ, BJ2, CD, TM, DA, HA, JCT, KC, Rahmani:2018hgp, DY, JAB, Wald:1977up, Wald:1978pj, Adler:1976jx,Adler:1977ac, Brown:1983zy}
with positive self-ajoint extensions of $A$ under the proper boundary conditions.
The renormalization  on the energy momentum tensor has already been intensively investigated in literature \cite{JCT, AS, BF, BF2, CD, KC, Rahmani:2018hgp, DY, JAB, TM, WOS:000497981800002, WOS:000488816700012, WOS:000471985900005, WOS:000471662400005, WOS:000469329900025, WOS:000450235800005, WOS:000450550300008, WOS:000441068300015, WOS:000441068300002, WOS:000439437600004, WOS:000437834000014, WOS:000432961800011, WOS:000429735200001, WOS:000424006300003, WOS:000416238500012, WOS:000404534100003, WOS:000423868800005, WOS:000391016900016, WOS:000415935400008, WOS:000388825000011, WOS:000382001700003, WOS:000378821700005, WOS:000361562900084, WOS:000355094700006, Namasivayam:2025dub}. For general spacetime, due to  the Hilbert space and vacuum not uniquely determined, one have difficulty to specify the  concrete physical renormalization condition. This shares some similarities as the case lack of asymptotic state  in Minkowski spacetime for dynamics  with asymptotic freedom.
Here for the specified background AdS spacetime,
we can fix the renormalization condition e.g.,  Eq. \eqref{RP}, (see also \cite{JCT, Thompson:2025jkn, Thompson:2025kfm}).
In the above we have shown the renormalization condition Eq. \eqref{RP}  removes the divergences not only for vacuum state, but also for
any excited states. Vacuum keeps as the self-consistent maximally symmetric solution of the
homogenous Einstein equations and  the
covariant conservation Eq. \eqref{nablatmunu} keeps. 
These are not dependent on
any boundary condition, which is the consequence of the quantization we employed here. Once the constant of the Lagrangian density is renormalized as $\Lambda_R=\Lambda_0-\delta \Lambda=\Lambda$, to get $\Lambda g_{\mu\nu}$ term in Einstein equations,
the AdS (background geometry) is the solution of the classical homogeneous Einstein Equations. Taking into account the fact that one can quantize and renormalize in various way, even the free field can give different EQFT after quantization. 

The expectation values of $T^{\mu\nu}_R$ on thermal and excited states may break the maximal symmetry, which can lead modifications to background geometry. One of the most used scheme to study the backreaction is the reparameterization of AdS metric on which quantization of matter fields is performed, then one can get the solution of the metric and the configurations of fields \cite{Allen:1986ty, JCT, Thompson:2025kfm, Diez-Tejedor:2011plw}. The other way is to study the evolution of Einstein-scalar system by parameterizing the spacetime and specifying initial data as the anti-de Sitter spacetime along with the expectation value of $T_R^{\mu\nu}$ \cite{PB}. 
However, the back-reactions from all of the states in the global Hilbert space as is well defined in the above have not been obtained. Without such an study, one can not get the conclusion on the (in)stability of a certain quantum dynamics in AdS. 

Perhaps the simplest way by only taking into account the 'minimal' quantum effects is to employ the coherent states. For example, if we consider the scalar field, the coherent states can be expressed  $a_{\omega j m}\left|\{\alpha_{\omega j m}\} \right\rangle =\alpha_{\omega j m}\left|\{\alpha_{\omega j m}\} \right\rangle ,\:\forall \omega \ge 0,j,m$, which are an overcomplete set of the global Hilbert space, hence any states in Hilbert can be expressed by 
$$\left| \Phi\right\rangle =\int \left[\prod_{\omega\ge 0 j m}d^2\alpha_{\{\omega j m\}}\right] c_{\alpha_{\{\omega j m\}}}\left|\{\alpha_{\omega j m}\} \right\rangle .$$
We define the semiclassical field $\eta(x)=\left\langle \Phi\right| \tilde{\phi}(x)\left| \Phi\right\rangle $ which satisfies the equation of motion 
$
	\hat{F}[\eta(x)]=-\left\langle \Phi|J(x)|\Phi\right\rangle , 
$Then one can use this semiclassical field $\eta(x)$ to calculate the semiclassical energy-momentum tensor $ T^{\mu\nu} _{\text{coh}}$. In this way, the method of \cite{PB} can be used here. In such consideration, the result had been obtained for the free massless scalar field theory with minimal coupling. This method is different from adopting matter sources as $\left\langle \Phi|T^{\mu\nu}_R|\Phi\right\rangle $ that involves more quantum effects than $T_{\text{coh}}^{\mu\nu}$. A full treatment of backreaction for general quantum states is technically involved and requiring dedicated analysis and we therefore defer it to future studies.

\section*{Acknowledgments}
This work is supported in part by National Natural Science Foundation of China (grant No. 12275157, 11775130, 11635009).
\appendix
\section{The calculation of renormalization counter-terms}
\subsection{The renormalization counter-terms of canonical energy-momentum tensor}
\label{appendix}
The vacuum state is utilized in the main text, we also can introduce 'general' degenerated vacuum states $$\left| \Psi \right\rangle  =\sum_ic_i\left| \psi_i\right\rangle, $$
which include any possible states relevant of ghost particles. Here
\begin{equation}\label{fock}
	\left| \psi_i \right \rangle =\prod_{\omega j m }	(B^\dagger_{\omega j m})^{{n'}^i_{\omega j m}}\prod_{\omega j m }	(A^\dagger_{\omega j m})^{n^i_{\omega j m}} \left| 0\right\rangle,\:n_{\omega j m}=0\:\mathrm{or}\:1.
\end{equation}
The product is for any number of values of $\omega<0$ and $j,m> 0$. For any linear operators, the conclusions can be straightforwardly applied to the linear combination of $\left| \psi_i\right\rangle $, $\left| \Psi\right\rangle $, so we will only concern on $\left| \psi_i\right\rangle $ and omit the subscript $i$. It is easy to show the 'general' degenerated vacua give the same results. 

On the one hand, it is straightforward to show that the Hamiltonian takes the same value for the degenerate vacua $\left| \psi\right\rangle $ and the vacuum (0-ghost) state $\left| 0\right\rangle $. On the other hand, in the following, we only focus on the expectation value of $T^{\mu\nu}$. 

Applying the expansion of $\phi$ Eq. \eqref{expansion4} and \eqref{h} to Eq. \eqref{TMUNU}, we could obtain
\begin{equation}\label{emtensor}
	\begin{split}
		&T^{\mu\nu}=\int_0^{+\infty}\int_0^{+\infty}\frac{d\omega d\omega^\prime   l}{\sqrt{2\omega2\omega^\prime}}\sum_{jj^\prime mm^\prime}\left[\left(f^{\mu\nu (0\circ)(0\circ)}_{\omega j m\omega^\prime j^\prime m^\prime}\right)a_{\omega j m}a_{\omega^\prime j^\prime m^\prime}+\left(f^{\mu\nu (0*)(0*)}_{\omega j m\omega^\prime j^\prime m^\prime}\right)a_{\omega j m}^\dagger a_{\omega^\prime j^\prime m^\prime}^\dagger\right]\\
		&\:\:\:\:\:\:\:\:+\int_0^{+\infty}\int_0^{+\infty}\frac{d\omega d\omega^\prime   l}{\sqrt{2\omega2\omega^\prime}}\sum_{jj^\prime mm^\prime}\left[\left(f^{\mu\nu  (0*)(0\circ)}_{\omega j m\omega^\prime j^\prime m^\prime}\right)a_{\omega j m}^\dagger a_{\omega^\prime j^\prime m^\prime}+\left(f^{\mu\nu  (0\circ)(0*)}_{\omega j m\omega^\prime j^\prime m^\prime}\right)a_{\omega j m} a_{\omega^\prime j^\prime m^\prime}^\dagger \right]\\
		&\:\:\:\:\:\:\:\:+\int_{-\infty}^{0}\int_{-\infty}^{0}\frac{d\omega d\omega^\prime   l}{\sqrt{2\omega2\omega^\prime}}\sum_{jj^\prime mm^\prime}^{m,m^\prime > 0}\left[\left(f^{\mu\nu (1\circ)(1\circ)}_{\omega j m\omega^\prime j^\prime m^\prime}\right)A_{\omega j m}A_{\omega^\prime j^\prime m^\prime}+\left(f^{\mu\nu (1*)(1*)}_{\omega j m\omega^\prime j^\prime m^\prime}\right)A_{\omega j m}^\dagger A_{\omega^\prime j^\prime m^\prime}^\dagger\right]\\
		&\:\:\:\:\:\:\:\:+\int_{-\infty}^{0}\int_{-\infty}^{0}\frac{d\omega d\omega^\prime   l}{\sqrt{2\omega2\omega^\prime}}\sum_{jj^\prime mm^\prime}^{m,m^\prime > 0}\left[\left(f^{\mu\nu (1*)(1\circ)}_{\omega j m\omega^\prime j^\prime m^\prime}\right)A_{\omega j m}^\dagger A_{\omega^\prime j^\prime m^\prime}+\left(f^{\mu\nu (1\circ)(1*)}_{\omega j m\omega^\prime j^\prime m^\prime}\right)A_{\omega j m} A_{\omega^\prime j^\prime m^\prime}^\dagger\right]\\
		&\:\:\:\:\:\:\:\:+\int_{-\infty}^{0}\int_{-\infty}^{0}\frac{d\omega d\omega^\prime   l}{\sqrt{2\omega2\omega^\prime}}\sum_{jj^\prime mm^\prime}^{m,m^\prime > 0}\left[\left(f^{\mu\nu (2\circ)(2\circ)}_{\omega j m\omega^\prime j^\prime m^\prime}\right)B_{\omega j m}B_{\omega^\prime j^\prime m^\prime}+\left(f^{\mu\nu (2*)(2*)}_{\omega j m\omega^\prime j^\prime m^\prime}\right)B_{\omega j m}^\dagger B_{\omega^\prime j^\prime m^\prime}^\dagger\right]\\
		&\:\:\:\:\:\:\:\:+\int_{-\infty}^{0}\int_{-\infty}^{0}\frac{d\omega d\omega^\prime   l}{\sqrt{2\omega2\omega^\prime}}\sum_{jj^\prime mm^\prime}^{m,m^\prime > 0}\left[\left(f^{\mu\nu (2*)(2\circ)}_{\omega j m\omega^\prime j^\prime m^\prime}\right)B_{\omega j m}^\dagger B_{\omega^\prime j^\prime m^\prime}+\left(f^{\mu\nu (2\circ)(2*)}_{\omega j m\omega^\prime j^\prime m^\prime}\right)B_{\omega j m} B_{\omega^\prime j^\prime m^\prime}^\dagger\right]\\
		&\:\:\:\:\:\:\:\:+\mathrm{mixing\:terms}, 
	\end{split}
\end{equation}
where the mixing terms include the mixing between positive and negative frequency terms. The definition of $f^{\mu\nu (i\circ)(k\circ)}_{\omega j m \omega^\prime j^\prime m^\prime}$ is\footnote{For example,
	$$
	f^{\mu\nu (0*)(0\circ)}_{\omega j m \omega^\prime j^\prime m^\prime}=\frac{1}{2}\left(g^{\mu\rho}g^{\nu\sigma}+g^{\nu\rho}g^{\mu\sigma}\right)\partial_\rho\phi_{\omega j m}^{0*}\partial_\sigma\phi_{\omega^\prime j^\prime m^\prime}^0-\frac{g^{\mu\nu}}{2}\left(g^{\rho\sigma}\partial_\rho\phi_{\omega j m}^{0*}\partial_\sigma\phi_{\omega^\prime j^\prime m^\prime}^0+\xi R\phi_{\omega j m}^{0*}\phi_{\omega^\prime j^\prime m^\prime}^0\right). 
	$$} 
$$\frac{1}{2}\left(g^{\mu\rho}g^{\nu\sigma}+g^{\nu\rho}g^{\mu\sigma}\right)\partial_\rho\phi_{\omega j m}^i\partial_\sigma\phi_{\omega^\prime j^\prime m^\prime}^k-\frac{g^{\mu\nu}}{2}\left(g^{\rho\sigma}\partial_\rho\phi_{\omega j m}^i\partial_\sigma\phi_{\omega^\prime j^\prime m^\prime}^k+\xi R\phi_{\omega j m}^i\phi_{\omega^\prime j^\prime m^\prime}^k\right), $$ where
\begin{equation*}
	\phi^i_{\omega j m}(x)=\left\{\begin{matrix}
		e^{-i\omega t}u_{\omega j}(r)Y_{jm}(\theta,\varphi)& i=0\\
		\cosh\omega tu_{\omega j}(r)Y_{jm}(\theta,\varphi)& i=1\\
		\sinh \omega t u_{\omega j}(r)Y_{jm}(\theta,\varphi)&i=2. 
	\end{matrix}\right.
\end{equation*}

It suffices to consider the example
\begin{equation}\label{example}
	\left| \psi\right\rangle =\left|\underbrace{0}_{\text{Boson Sector}};\underbrace{0_A^{\omega_1<0,j_1,m_1\ge 1},1_A^{\omega_2<0,j_2,m_2\ge 1}\cdots ;0_B^{\omega_3<0,j_3,m_3\ge 1},1_B^{\omega_4<0,j_4,m_4\ge 1},\cdots}_{\text{Ghost Sector}}\right\rangle, 
\end{equation}
where ellipsis refers to possible ghost modes containing one and zero ghost particle. 

The expectation of the mixing terms for Eq. \eqref{example} apparently vanishes. Then the expectation of the first four terms of Eq. \eqref{emtensor} is
$$\int_0^{+\infty}d\omega\frac{l}{2\omega}\sum_{j=0}^{+\infty}\sum_{m=-j}^jf^{\mu\nu(0\circ)(0*)}_{\omega j m\omega j m}. $$
It is easy to check the expectations of the 5th, 6th, 9th and 10th terms vanish. Moreover, we obtain the summation of the expectation of 7th and 8th terms of Eq. \eqref{emtensor} is 
\begin{equation*}
	\begin{split}
		\int_{-\infty}^0\frac{d\omega l}{-2\omega}\sum_{j=1}^{+\infty}\sum_{m>0}f^{\mu\nu(1\circ)(1*)}_{\omega j m\omega j m}, 
	\end{split}
\end{equation*}
with the help of Eq. \eqref{aca}. 
Similarly, the expectation of 11th and 12th term is
\begin{equation}
	\int_{-\infty}^0\frac{d\omega l}{-2\omega}\sum_{j=1}^{+\infty}\sum_{m>0}f^{\mu\nu(2\circ)(2*)}_{\omega j m\omega j m}. 
\end{equation}
As a consequence, the expectation of energy-momentum tensor Eq. \eqref{emtensor} in vacuum state $\left|\psi\right\rangle$ is obtained by summing the contributions from 1st to 12th term, namely
\begin{equation}\label{expT}
	\left\langle \psi|T^{\mu\nu}|\psi\right\rangle=\int_0^{+\infty}d\omega\frac{l}{2\omega}\sum_{j=0}^{+\infty}\sum_{m=-j}^jf^{\mu\nu(0\circ)(0*)}_{\omega j m\omega j m}+\sum_{i=1,2}\int_{-\infty}^0\frac{d\omega l}{-2\omega}\sum_{j=1}^{+\infty}\sum_{m>0}f^{\mu\nu(i\circ)(i*)}_{\omega j m\omega j m},
\end{equation}
where $\left| \psi\right\rangle $ is involved with the boson vacuum and arbitrary mutiple ghost particles. 

\subsection{The renormalization counter-terms of metric energy-momentum tensor}
\label{improvememt}

The energy-momentum tensor taking into account the derivative of $R$ with respective of $g_{\mu\nu}$ is 
\begin{equation}
	\tilde{T}^{\mu\nu}=T^{\mu\nu}_{can}+\xi\left[g^{\mu\nu}\Box \phi^2-D^\mu D^\nu\phi^2+R^{\mu\nu}\phi^2\right], 
\end{equation}
corresponding Hamiltonian is
\begin{equation}
	\tilde{H}=H+\xi \int dr d\theta d\varphi \partial_r\left[\sin\theta \left(-r^2 (1+l^{-2}r^2) \partial_r \phi^2\right)+r^3 l^{-2}\phi^2\right] . 
\end{equation}
Hence, $\tilde{H}$ and $H$ only differ up to a spacial boundary term \cite{BF, BF2}. As usual, we only study quantization inside the AdS spacetime $r<\infty$, so the quantization is exactly the same. It is clear $\tilde{T}^{\mu\nu}$ and $T^{\mu\nu}_{can}$ are indeed different, hence we should introduce the improved counter-terms for satisfying the physical renormalization condition $\left\langle 0\right| \tilde{T}^{\mu\nu}_R\left| 0\right\rangle =0$ as $T_{can}^{\mu\nu}$, 
\begin{equation}\label{counterterms3}
	\begin{split}
		&\Delta \tilde{T}^{\mu\nu}=\Delta \tilde{T}^{\mu\nu}_1+\Delta \tilde{T}^{\mu\nu}_2 \\
		&\Delta \tilde{T}^{\mu\nu}_1=-\int_0^\infty d\omega\sum_{j,m}\frac{l}{2\omega}\tilde{f}^{\mu\nu(0\circ)(0*)}_{\omega j m\omega j}\\
		&\Delta \tilde{T}^{\mu\nu}_2=\sum_{i=1,2}\int_{-\infty}^0\frac{d\omega l}{2\omega}\sum_{j=1}^{+\infty}\sum_{m>0}\tilde{f}^{\mu\nu(i\circ)(i*)}_{\omega j m\omega j m},
	\end{split}
\end{equation}
where $$\tilde{f}^{\mu\nu (i\circ)(k\circ)}_{\omega j m \omega^\prime j^\prime m^\prime}=f^{\mu\nu (i\circ)(k\circ)}_{\omega j m \omega^\prime j^\prime m^\prime}+\xi \left[g^{\mu\nu}\Box \phi^i_{\omega j m}\phi^k_{\omega' j' m'}-g^{\mu\rho}g^{\nu\sigma}D_\rho D_\sigma\phi^i_{\omega j m}\phi^k_{\omega' j' m'}+R^{\mu\nu}\phi^i_{\omega j m}\phi^k_{\omega' j' m'}\right]. $$
Then, taking the expectation of physical states \eqref{excited} on $\tilde{T}_R^{\mu\nu}=\tilde{T}^{\mu\nu}+\Delta \tilde{T}^{\mu\nu}$, we can obtain a different value $\left\langle\tilde{T}_R^{\mu\nu} \right\rangle $ remaining covariant conserved $D_\mu \left\langle \tilde{T}_R^{\mu\nu}\right\rangle =0$. 

\section{The Green Functions in the mode function space}
\label{c}
From the Yang-Feldman equation, $\tilde{\phi}$ keeps in the mode function space free from the negative degrees of freedom provided that the Green function is well defined in the mode function space free from the negative degrees of freedom. For this sake, we expand the Green function by all the mode functions $\phi_{\omega j m\:}(\omega \ge 0)$ 
\begin{equation}
	G(x,y)=\int_{-\infty}^{+\infty} \frac{dp_0}{2\pi}\int_0^\infty d\omega \sum_{j=0}^{+\infty}\sum_{m=-j}^{j}\tilde{G}(p_0,\omega,j,m)f_\omega^2 e^{-ip_0(x^0-y^0)}\phi_{\omega j m}(r_x,\theta_x,\varphi_x)\phi^*_{\omega j m}(r_y,\theta_y,\varphi_y). 
\end{equation}
With the help of the following equation
\begin{equation}
	\delta^{4}(x,y)=\frac{l}{1+l^{-2}r^2}\int_{-\infty}^{+\infty} \frac{dp_0}{2\pi}\int_0^\infty d\omega \sum_{j=0}^{+\infty}\sum_{m=-j}^{j}f_\omega^2 e^{-ip_0(x^0-y^0)}\phi_{\omega j m}(r_x,\theta_x,\varphi_x)\phi^*_{\omega j m}(r_y,\theta_y,\varphi_y), 
\end{equation}
which is consistent with complete relation Eq. \eqref{wb}, then employing the equation of motion $\hat{F}_x G(x,y)=\delta^4(x,y)$, we have 
\begin{equation}
	\begin{split}
		&G(x,y)=\int_{C}\frac{dp_0}{2\pi}\int _0^\infty d\omega \sum_{j=0}^{+\infty}\sum_{m=-j}^j\frac{l}{(p_0-\omega)(p_0+\omega)}f_\omega^2e^{-ip_0(x^0-y^0)}\phi_{\omega j m}(r_x,\theta_x,\varphi_x)\phi^*_{\omega j m}(r_y,\theta_y,\varphi_y), 
	\end{split}
\end{equation}
it is obviously that the Green function in the mode function space, provided that the integration contour $C$ can be deformed from the isolated singularities, which stems from the second order temporal derivative Eq. \eqref{eom1}. The special contour determines the specific boundary condition. 

The general time-ordered correlation function is $\langle \Omega|T[\tilde{\phi}(x)\tilde{\phi}(x')]|\Omega\rangle$, here the interacting vacuum $\left| \Omega\right\rangle $ satisfies $H_{\tilde{\phi}}\left|\Omega\right\rangle =0$ with $H_{\tilde{\phi}}$ the full Hamiltonian. This correlation functions can be expressed by the Casimir spectrum representation (corresponding to Källén–Lehmann spectral representation​ in Minkowski spacetime). Since, the interacting field $\tilde{\phi}$ is well defined within the mode function space, so the specturm representation is free from the poles of negative degrees of freedom.

\bibliographystyle{unsrt}
\bibliography{ref.bib}

\end{document}